\documentclass[5p,twocolumn]{elsarticle}
\usepackage{graphicx,latexsym}
\usepackage{dcolumn}
\usepackage{amssymb,amsmath,bm}
\usepackage{subfigure}
\usepackage{braket}
\usepackage{siunitx}
\usepackage{array}
\usepackage{float}
\usepackage[nopar]{lipsum}
\usepackage{caption}
\usepackage{multirow}
\usepackage{bigstrut}

\biboptions{sort&compress}

\usepackage[super]{nth}

\newcommand{\angstrom}{\text{\normalfont\AA}}
\usepackage{hyperref}
\hypersetup{
    pdfnewwindow=true,       
    colorlinks=true,         
    linkcolor=blue,          
    citecolor=blue,          
    filecolor=magenta,       
    urlcolor=black           
}

\usepackage[normalem]{ulem}

\def\sec#1{Sec.\ \ref{#1}}

\def\fig#1{Fig.\ \ref{#1}}
\def\tab#1{Tab.\ \ref{#1}}

\journal{}

\begin{document}

\begin{frontmatter}


\title{Planar buckling controlled optical conductivity of SiC monolayer from Deep-UV to visible light region: A first-principles study}

\author[a1,a2]{Nzar Rauf Abdullah}
\ead{nzar.r.abdullah@gmail.com}
\address[a1]{Division of Computational Nanoscience, Physics Department, College of Science,
             \\ University of Sulaimani, Sulaimani 46001, Kurdistan Region, Iraq}
\address[a2]{Computer Engineering Department, College of Engineering,
	\\ Komar University of Science and Technology, Sulaimani 46001, Kurdistan Region, Iraq}
\author[a1]{Hunar Omar Rashid}
\author[a3]{Botan Jawdat Abdullah}
\address[a3]{Physics Department, College of Science- Salahaddin University-Erbil, Erbil 44001, Kurdistan Region, Iraq}

\author[a4]{Chi-Shung Tang}
\address[a4]{Department of Mechanical Engineering,
	National United University, 1, Lienda, Miaoli 36003, Taiwan}

\author[a5]{Vidar Gudmundsson}
\address[a5]{Science Institute, University of Iceland, Dunhaga 3, IS-107 Reykjavik, Iceland}


\begin{abstract}

The electrical and optical properties of flat and planar buckled siligraphene (SiC) monolayer are examined using a first principles approach. Buckling between the Si and the C atoms in SiC structures influences and impacts the properties of the 2D nanomaterial, according to our results. The electron density of a planar SiC monolayer is calculated, as well as the effects of buckling on it. According to our findings, a siligraphene monolayer is a semiconductor nanomaterial with a direct electronic band gap that decreases as the planar buckling rises. The contributions to the density of states differ owing to changes in the system's structure. Another explanation is that planar buckling reduces the sp$^2$ overlapping, breaking the bond symmetry causing it to become a sp$^3$ bond.
We show that increased planar buckling between the Si and the C atoms alters the monolayer's optical, mechanical, and thermal properties. A managed planar buckling increases the optical conductivity with a significant shift in the far visible range, as all optical spectra features are red shifted, still remaining visible. Instead of a $\sigma\text{-}\sigma$ covalent bond, the sp$^3$ hybridization produces a stronger $\sigma\text{-}\pi$ bond. Optical characteristics such as the dielectric function, the absorbance, and the optical conductivity of a SiC monolayer are investigated for both parallel and perpendicular polarization of the incoming electric field for both flat and planar buckled systems. The findings show that the optical properties are influenced for both of these two polarizations, with a significant change in the optical spectrum from the near visible to the far visible. The ability to manipulate the optical and electrical characteristics of this critical 2D material through planar buckling opens up new technological possibilities, especially for optoelectronic devices.

\end{abstract}


\end{frontmatter}

\section{Introduction}

Nanomaterials are important in the development of a wide range of modified-material devices, and their high specific surface areas and unique physical properties will make them useful in application. Two-dimensional (2D) monolayers are important in today's research due to their improved magnetic, electrical, optical, mechanical, and catalytic properties compared to their bulk form \cite{ gupta2015recent, peng2015two, abdullah2021properties}. One of the most essential characteristics that distinguishes nanomaterials from other materials is their dimensionality. Graphene \cite{novoselov2004electric}, the first 2D monolayer, was discovered in 2004, but other materials have also been synthesized experimentally \cite{mak2016photonics, vogt2012silicene, mannix2018borophene, zhu2015epitaxial}.

Novel 2D materials with modest band gaps, high carrier mobilities, and great stability are highly wanted for applications. Among them, the siligraphene monolayer, a 2D form of silicon carbide, is one of the most fascinating SiC, that can be used in a range of applications. Carbon, the main component of graphene, tends to form sp$^2$ hybridization in 2D structures, which are plentiful and safe on Earth. Silicon, on the other hand, is the semiconductor industry's present basis. Both components are low-cost resources, that are an ideal options for building optical applications. Siligraphene monolayers with varying Si/C concentrations have surpassed pure graphene and silicene with improved mechanical, electrical, optical, and thermodynamic properties \cite{zheng2019adsorption, anikina2020elucidating, houmad2018thermal, guan2019computational, houmad2016optical, abdullah2020electronic}.

Many attempts have been made to experimentally synthesize siligraphene monolayers, including nanocleavage exfoliation in polar solutions like Nmethylpyrrolidone and isopropyl alcohol to reduce SiC to a single layer or few layers with thicknesses as low as 0.5-1.5 nm, according to findings reported in the literature. These ultrathin freestanding SiC sheets, which are made composed of graphitic SiC, graphene, and embedded ultrathin wurtzite SiC, have excellent photoluminescence properties \cite{lin2012light}. In another study, distinct 2D SiC nanosheets with an average interlayer distance of 0.255 nm were created by combining a carbothermal reaction with a post-sonication phase. The average thickness of the free-standing nanosheets was found to be $2\text{-}3$ nm \cite{chabi2016graphene}. A reactive sintering technology based on a two-step sintering procedure in the [111] direction was successfully used to generate SiC nanosheets with an average thickness of 33.7 nm on original graphite surfaces \cite{chen2019low}, and a wet exfoliation process was applied for the first time in 2021 to successfully separate 2D SiC from hexagonal SiC. Unlike many other 2D materials, such as silicene, which tends to deteriorate, the 2D SiC nanosheets developed are stable and exhibit no signs of degradation. According to a study, the fabricated 2D SiC also generates visible light \cite{chabi2021creation}.

The ability to fine-tune the material properties of a monolayer opens the door to a wide range of applications. Many approaches, including the use of doping, strain, electric field, and the tuning of planar buckling, have been proposed theoretically to effectively control their essential properties. Doping is a way for fine-tuning a material's electronic structure and other characteristics. The dopants are transition metals (Ti, V, Cr, Mn, and Fe) in the SiC monolayers that have been studied using density functional theory to investigate their structural, electrical, and magnetic properties. These dopants changed the host's electrical and magnetic properties by generating gap states inside the SiC's band structure \cite{majid2020first}. The electrical structure and the optical properties of rare earth atoms doped in SiC monolayers, such as La, Ce, and Th, are investigated using the first-principle approach. Pure 2D SiC is a direct-gap semiconductor with a gap of 2.60 eV. The La-dopeding of a 2D SiC introduces an impurity states at the Fermi energy, and the band gap is 1.93 eV, after a doping where Si atoms are replaced with rare earth atoms. They become quasi-direct band-gap semiconductors when Ce or Th is added to them. Doping with La increases the static dielectric constant of 2D SiC, whereas Th has minimal effects \cite{wan2019photoelectric}.

The characteristics of a 2D monolayer can be influenced by both strain and electric fields. Density functional theory calculations have been employed to investigate the structural, the electrical, and the optical characteristics of a layered SiC monolayer, that has been modified for in-plane or out-of-plane strain effects. Strain may be used to generate a tunable band structure and optical properties, such as indirect-direct band gap transitions, band gap shrinking and enlargement, and red or blue shifts of absorption peaks \cite{xu2016controlling, ABDULLAH2022115554}. The structure of SiC monolayers under unidirectional strain regimes has been explored using density functional calculations. The material deforms anisotropically in the stress domain, exhibiting nonlinear elastic deformation. Under compressive 10.35 percent strain, the indirect band gap of a 2D SiC monolayer turns to a direct band gap, and at 16.87 percent, it becomes a metal. In addition, structural-electronic-bonding findings show that the design of a band gap may be controlled by using a unidirectional strain regime to reduce the band gap \cite{belarouci2018two}.

On the other hand, the gas-sensing capabilities of SiC monolayer can be improved by functionalizing the materials, such as by adding an external electric field. According to first-principles considerations, the SiC$_7$ monolayer is optimal for use as a gas sensing material for detecting NO, NO$_2$, and NH$_3$. Furthermore, the applied electric field can change the electrical and optical characteristics of molecules adsorbed on the SiC$_7$ monolayer \cite{zhao2021enhancement}, and The effect of hydrogen adsorption on the SiC monolayer was studied using density-functional theory approach, which took into account several adsorption process configurations. Despite having almost comparable electrical characteristics, the chair-like shape is found to be more stable, with a binding energy of roughly -3.845 eV for hydrogen atoms. Owing to hydrogen adsorption the plasmon frequency of the chair-like, hydrogenated monolayer is shifted and the optical characteristics changed producing a drop in the dielectric constant and the static refractive index \cite{delavari2018electronic}.

Another fascinating way of modifying physical features is a planar buckling modification. Buckling effects on the electrical and optical characteristics of beryllium oxide monolayers were previously investigated using a first-principles approach with moderate planar buckling. Planar buckling can lead to a variation of the band gap, according to electrical investigations, since the energy band gap reduces when the planar buckling parameter is increased. Furthermore, increasing the planar buckling factor leads to red-shift of all optical spectra features to lower energy \cite{jalilian2016buckling}. The electrical and optical properties of a hexagonal boron nitride monolayer have been investigated using a density functional analysis to modify the planar buckling. As a result, the wide band gap of a flat BN monolayer is reduced to a narrower band gap in a buckled BN monolayer, improving optical activity in the Deep-UV region. Additionally, increased planar buckling enhances the optical conductivity in both the visible and the deep-UV domains, depending on the polarization orientation of the incoming light \cite{abdullah2022enhanced}.

In this work, the electrical and optical characteristics of SiC monolayers are investigated using first-principles calculations for flat and buckled structures with a range of planar buckling values. We find that planar buckling of SiC monolayers can generate a variable band gap, which decreases as the planar buckling increases. All related impact factors, such as the tuning of the density of state, the elongation of the bond length between the Si and C, and the change in the charge density distribution, are analyzed for structural modification. All key parameters such as the dielectric function, the absorption coefficient, and the optical conductivity change and are examined when the planar buckling factor is enhanced. Our findings give rise to new information about optoelectronic functionality in nanotechnology.

The computational methodologies and model structure are briefly discussed in \sec{Sec:Methodology}. The major achieved outcomes are examined in \sec{Sec:Results}. The conclusion of the results is reported in \sec{Sec:Conclusion}

\section{Methodology}\label{Sec:Methodology}

We consider a 2D SiC monolayer structure consisting of a $2\times 2$ supercell with equal number of Si and C atoms. The SiC monolayer is fully relaxed with the plane-waves kinetic energy and charge densities cutoff assumed to be $1088.5$~eV, and $1.088 \times 10^{4}$~eV, respectively. The SiC monolayer is relaxed until all the forces on the atoms are less than $10^{-5}$ eV/$\angstrom$ with a $18 \times 18 \times 1$ Monkhorst-Pack grid.
The interaction of individual SiC monolayers is canceled out by assuming a large wide vacuum layer of $20 \, \angstrom$ in the $z$-direction.
The generalized gradient approximation (GGA) with the Perdew-Burke-Ernzerhof (PBE) functionals is used to approximate the exchange and the correlation term \cite{PhysRevB.54.11169}. The GGA-PBE is implemented in the Quantum espresso (QE) software \cite{Giannozzi_2009, giannozzi2017advanced}. The DFT approach in the QE software is implemented based on the Kohn-Sham density functional theory (KS-DFT).
The band structure and the density of states (DOS) are obtained by utilizing the Self-Consistent Field (SCF) and non-self-consistent field (NSCF) calculations, respectively. In these calculations, we have used
a Monkhorst-Pack grid of $18 \times 18 \times 1$ for the SCF  and $100 \times 100 \times 1$ for the NSCF.
The optical properties of the SiC monolayers are obtained using QE with the optical broadening of $0.1$~eV.
In order to check the thermodynamic stability of the flat and planar SiC monolayer, ab-initio molecular dynamics (AIMD) calculations are done
for an NVT ensemble for $10$~ps with a time step of $1.0$~fs, using the Nosé-Hoover heat bath
approach described by Martyna and Klein \cite{doi:10.1063/1.463940}.

\section{Results}\label{Sec:Results}

\begin{figure}[htb]
	\centering
	\includegraphics[width=0.5\textwidth]{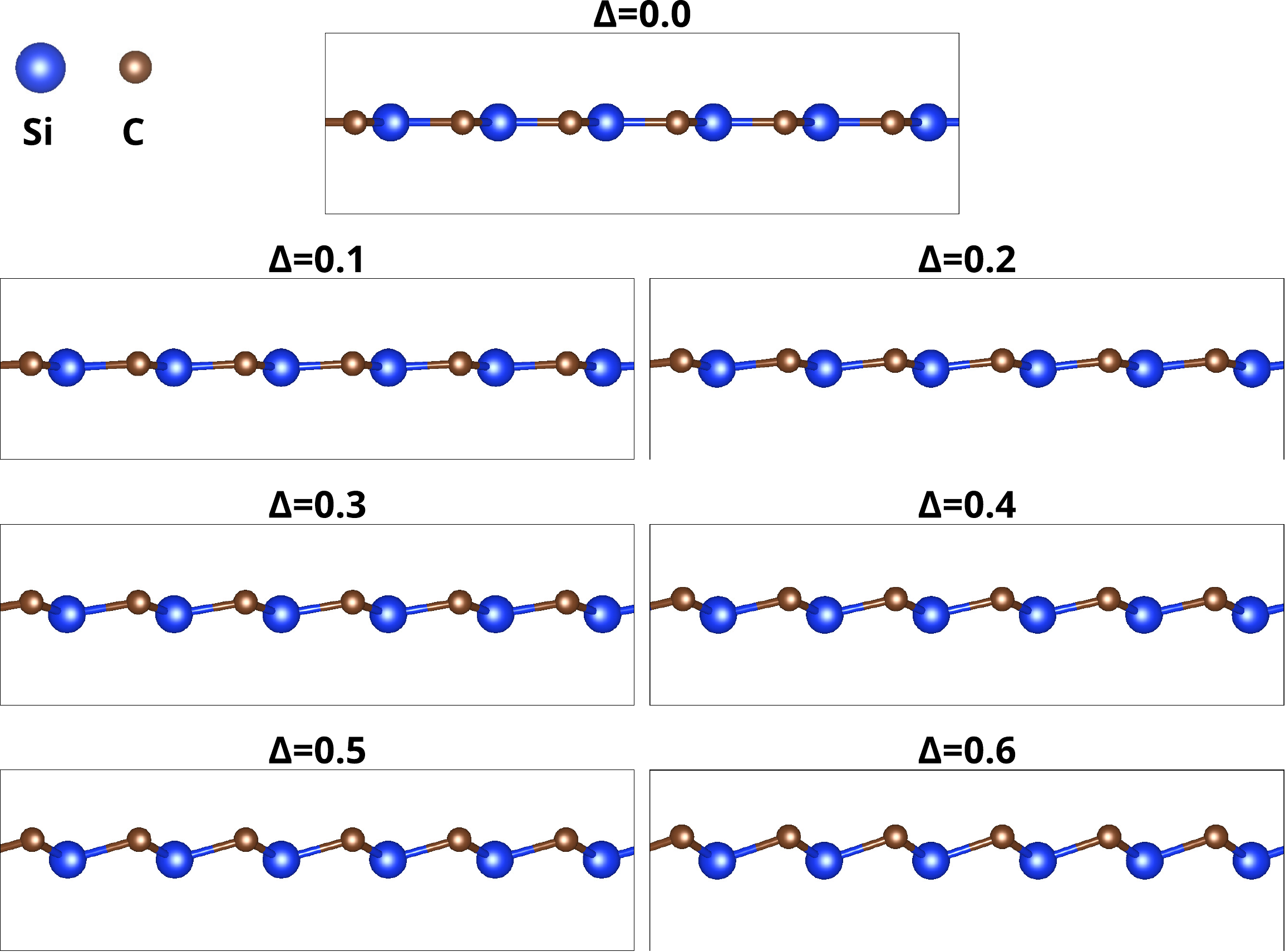}
	\caption{Side view of crystal structures of SiC monolayer for different values of $\Delta$.}
	\label{fig01}
\end{figure}

In this section, we demonstrate the electronic and optical properties of SiC monolayer for varying planar buckling strength, $\Delta$. The side view of crystal structures of SiC monolayer for different values of $\Delta$ is shown in \fig{fig01}, where the influecnes of $\Delta$ on the atomic positions are clearly seen. The SiC monolayer consists of a combination of carbon atoms known for forming graphene and silicon atoms known to form silicence. It is well known that graphene has a stable planar structure ($\Delta = 0.0$), while the planar buckling of silicene is $\Delta \approx 0.45$~$\angstrom$~\cite{ABDULLAH2021114644}. The combination of the C and the Si atoms creates a SiC monolayer with vanishing planar buckling $\Delta = 0.0$.
It has been shown that a 2D SiC monolayer has a flat shape with $100\%$ planar structure with inherent dynamic stability \cite{Chabi_2016}. The planar 2D SiC monolayer is thus energetically very stable.
Several studies have investigated the stability of a planar SiC monolayer, and all these studies have demonstrated that a 2D SiC is energetically stable and has an inherent dynamic stability \cite{nano10112226, doi:10.1021/jp210536m}.
The predicted planarity property of SiC monolayer is significant in contributing to the development of unprecedented characteristics. In addition, planar buckling plays an important role in controlling the physical properties of SiC monolayer.
Understanding the underlying mechanism of the buckling behavior in SiC monolayer is important for the design of siligraphene and its composites. We therefore explore the bucking effects and the mechanisms that modify the SiC monolayer due to the planar buckling \cite{ABDULLAH2022106943}.

\subsection{Electronic properties}

The 2D SiC monolayers have a graphene-like honeycomb structure consisting of alternating Si and C atoms. Similar to graphene, the C and the Si atoms form $sp^2$ hybrid orbital to create the SiC nanosheet.
The hybridization order is strongly affected by the value of $\Delta$, which causes an orbital redirection and bond reconstruction.

\begin{table}[h]
	\centering
	\begin{center}
		\caption{\label{table_one} Percentage of orbital contribution in bond character and hybridization bond order.}
		\begin{tabular}{l|l|l|l}\hline
			$\Delta$ ($\angstrom$)& $s$ ($\%$)  &  $p$ ($\%$) &  Hybrid      \\ \hline
			0.0	    & 33.33 & 66.66 &  sp$^2$      \\
			0.1	    & 33.10 & 66.90 &  sp$^{2.021}$ \\
			0.2	    & 32.47 & 67.53 &  sp$^{2.079}$ \\
			0.3	    & 31.44 & 68.56 &  sp$^{2.180}$ \\
			0.4	    & 29.97 & 70.03 &  sp$^{2.336}$ \\
			0.5	    & 28.07 & 71.93 &  sp$^{2.562}$ \\
			0.6	    & 25.77 & 74.23 &  sp$^{2.88}$ \\
			\hline
	\end{tabular}	\end{center}
\end{table}

In fact, the planar buckling influences the $\sigma$ and the $\pi$ bonds of the SiC monolayer leading to enhanced $\sigma\text{-}\pi$ common bonds.
The $sp^2$-hybridization is thus changed towards an $sp^3$-hybridization for increasing value of $\Delta$ as is presented in \tab{table_one}.
For instance, the hybridization becomes $sp^{2.88}$ at $\Delta = 0.6$~$\angstrom$, and the planar buckling thus reduces the $sp^2$ overlapping, and the bond symmetry is broken simultaneously.
The maximum allowed value of $\Delta$ is $0.6$~$\angstrom$ for the SiC monolayer, and
the hybridization would becomes $sp^{>3}$ if the value of the planar buckling
were further increased to $0.7$~$\angstrom$, which is an impossible state.
The $sp^3$ bonds instead of the usual $sp^2$ bonds are also found in SiC nanotube structures \cite{SHI2013319}. 

\begin{figure}[htb]
	\centering
	\includegraphics[width=0.44\textwidth]{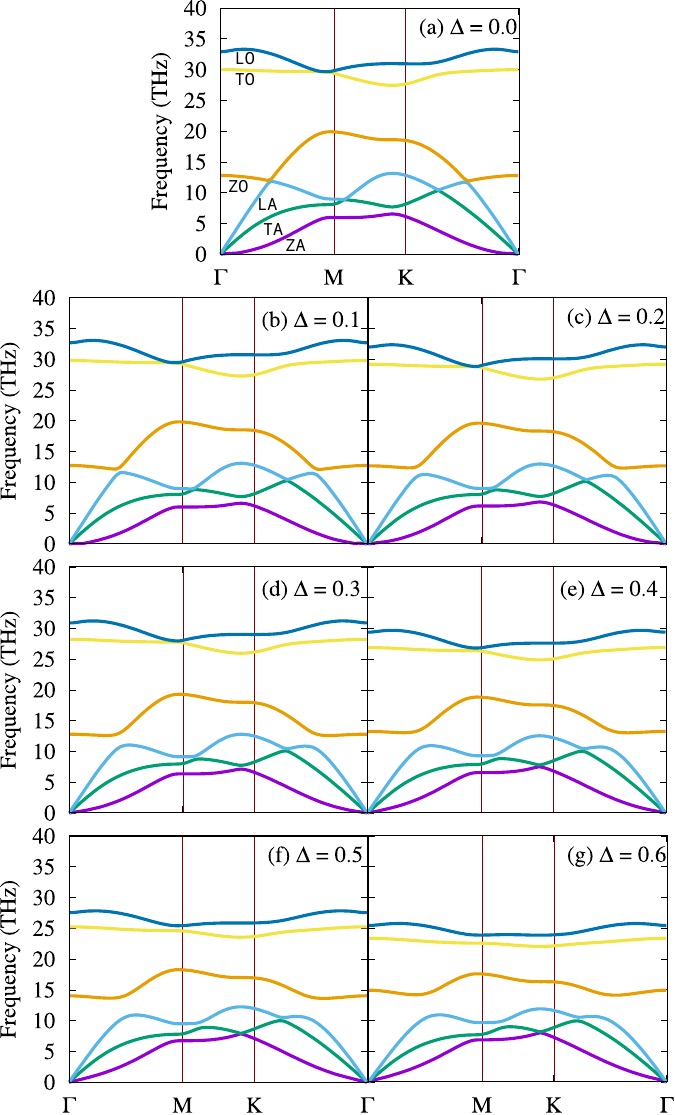}
	\caption{Phonon band structure for different values of $\Delta = 0.0$ (a), $0.1$ (b), $0.2$ (c), $0.3$ (d), $0.4$ (e), $0.5$ (f), and $0.6$~$\angstrom$ (g).}
	\label{fig02}
\end{figure}

The formation energy, the Si-C bond length, and the band gap are influenced by the planar buckling as
is presented in \tab{table_two} \cite{ABDULLAH2022106835}. The formation energy is the energy required for generating the configuration of a monolayer.

The formation energy of the buckled SiC monolayers is determined relative to a flat layer
	\begin{equation}\label{eq01}
		E_f = E_{\rm Buckling} - E_{\rm Flat},
	\end{equation}
where E$_{\rm Buckling}$ indicates the total energy of a SiC monolayer with planar buckling, and E$_{\rm Flat}$ defines the total energy of the flat SiC monolayer, without buckling.
The low formation energy is an indication of the high energetically stability of the SiC monolayer.
The lower the formation energy, the more stable structure is obtained. We can therefore see that the planar SiC monolayer is the most energetically stable structure, and the stability is slightly decreased with increasing $\Delta$.

\begin{table}[h]
	\centering
	\begin{center}
		\caption{\label{table_two} The formation energy (E$_{f}$), the bond length of Si-C, and the band gap (E$_g$) for different values of planer buckling, $\Delta$.}
		\begin{tabular}{l|l|l|l}\hline
	   $\Delta$	   &E$_f$ (eV/atoms)&  Si-C ($\angstrom$)      &  E$_g$ (eV)    \\ \hline
			0.0	   &    -           & 1.781                    &  2.524         \\
			0.1	   &   0.111        & 1.784                    &  2.452         \\
			0.2	   &   0.442        & 1.792                    &  2.260         \\
			0.3	   &   0.999        & 1.806                    &  2.014         \\
			0.4	   &   1.781        & 1.826                    &  1.755         \\
			0.5	   &   2.807        & 1.850                    &  1.514         \\
			0.6	   &   4.105        & 1.880                    &  1.305         \\ \hline
	\end{tabular}	\end{center}
\end{table}

The dynamical stability of an SiC monolayer can be tested using the phonon band structure shown in \fig{fig02}. The phonon band structure along the high symmetry directions is calculated. The phonon band structure of a flat SiC monolayer, $\Delta$ = 0.0 (a),
has no imaginary frequencies, displaying the dynamical stability of the monolayer.
The phonon modes are classified in to acoustic and optical modes.
The acoustic phonon modes are in-plane longitudinal acoustic (LA), in-plane transverse acoustic (TA), and out-of-plane acoustic (ZA). In the same way, the optical phonon modes are in-plane longitudinal optical (LO), in-plane transverse optical (TO), and out-of-plane optical (ZO) modes.
The phonon band structure of the buckled SiC monolayers does not display any imaginary frequencies showing the dynamical stability, even at high values of the planar buckling.

\begin{figure}[htb]
	\centering
	\includegraphics[width=0.42\textwidth]{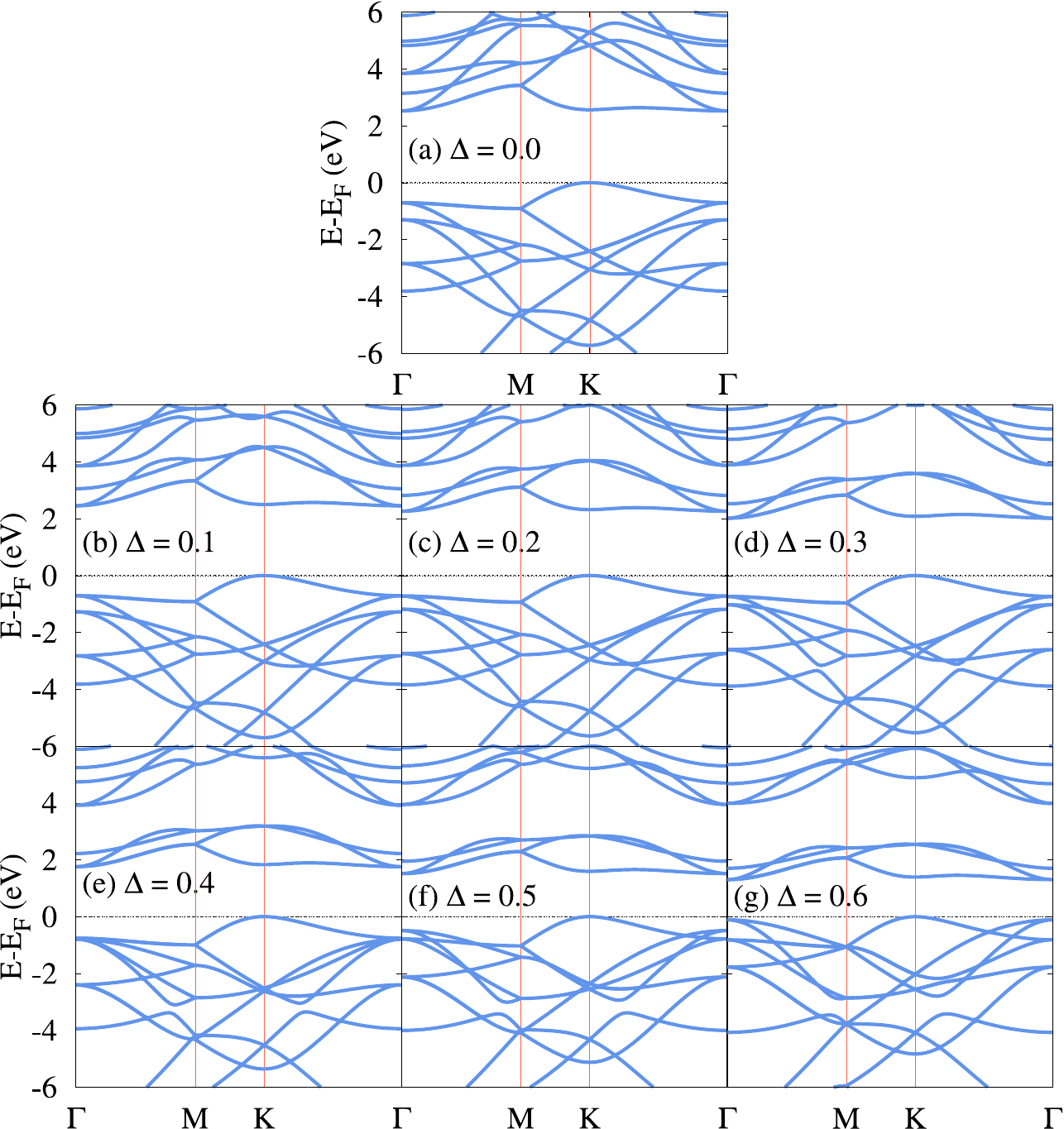}
	\caption{Band structure for optimized SiC monolayers with planer buckling, $\Delta = 0.0$ (a), $0.1$ (b), $0.2$ (c), $0.3$ (d), $0.4$ (e), $0.5$ (f), and $0.6$~$\angstrom$ (g).
		The energies are with respect to the Fermi level, and the Fermi energy is set to zero.}
	\label{fig03}
\end{figure}

The structure modification caused by the planar buckling changes the Si-C bond length, while the lattice constant is slightly changed (see \tab{table_two}).
The longer bond lengths indicate that the electrons are less tightly bound to the
atoms. In this case, less energy is required to remove an electron from an atom
leading to a decreased band gap \cite{gacevic:hal-00632224}. The longer bond lengths thus corresponds to smaller band gaps. We should expect that the SiC monolayer with $\Delta = 0.6$~$\angstrom$ has the smallest band gap among all the considered values of $\Delta$ (see \tab{table_two}).

The band structure of SiC monolayer shown in \fig{fig03} confirms the band gap reduction due to the planar buckling. Our results indicate that a flat SiC monolayer, $\Delta = 0.0$, is a direct band gap semiconductor in contrast to the indirect band gap in bulk SiC.
The formation of Dirac cones or the direct band gap of a flat SiC monolayer is attributed to the preservation of the $\pi$-conjugate orbitals and the hexagonal symmetry.
The calculated direct band gap of a flat SiC monolayer is $2.524$~eV using GGA-PBE, which is an underestimated value compared to the band gap of $3\text{-}4.8$~eV computed with the GW quasi-particle correction, GLLB-SC, method and other approximations \cite{Alaal_2016, PhysRevB.84.085404, PhysRevB.81.075433}.
Increasing the planar buckling the direct band gap is converted to an indirect band gap with a smaller value especially at $\Delta = 0.6$~$\angstrom$. The direct-indirect band gap transition can be referred to the destruction of the $\pi$-conjugate orbitals and the breaking of the hexagonal symmetry in the presence of planar buckling.
The direct-indirect band gap transition phenomena in 2D SiC is also seen in 2D transition metal dichalcogenides (TMDs) \cite{PhysRevB.80.155453}.

\begin{figure}[htb]
	\centering
	\includegraphics[width=0.35\textwidth]{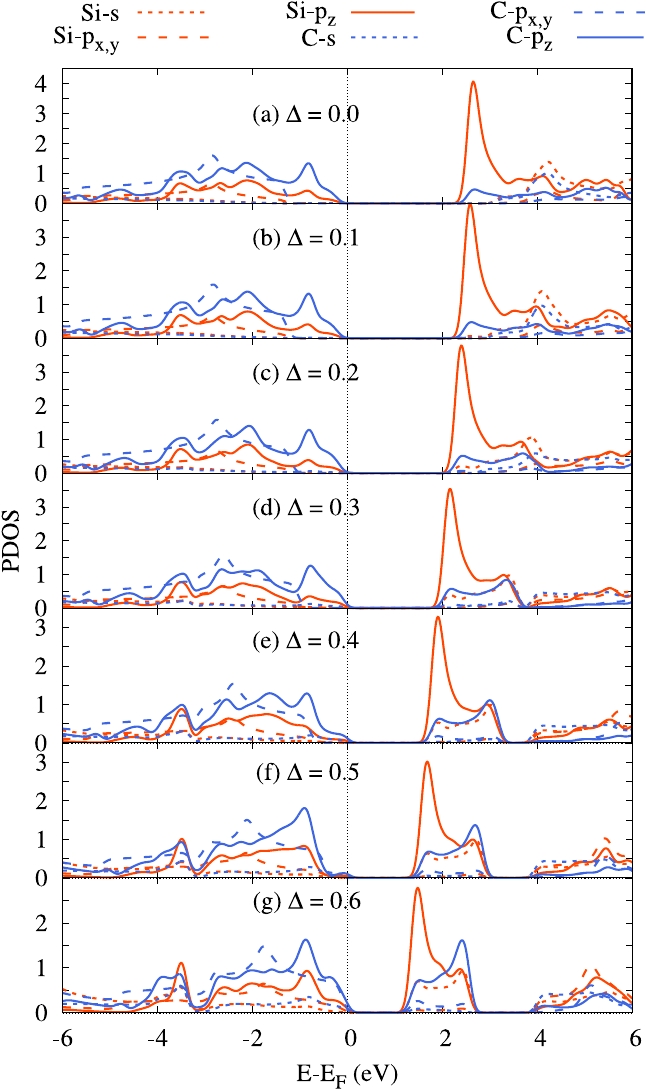}
	\caption{Partial density of states, PDOS of the SiC monolayers with planar buckling, $\Delta = 0.0$ (a), $0.1$ (b), $0.2$ (c), $0.3$ (d), $0.4$ (e), $0.5$ (f), and $0.6$~$\angstrom$ (g). The energies are with respect to the Fermi level, and the Fermi energy is set to zero.}
	\label{fig04}
\end{figure}

To get insight into the band gap reduction and the band structure modification due to the planar buckling, the partial density of states (PDOS) of a SiC monolayer for different values of $\Delta$ is presented in \fig{fig04}. It is obvious that the density of valence states near the Fermi energy is formed by a hybridization of the $p_z$-orbitals of Si and C atoms, where the $p_z$-orbital of the C atoms is dominant for a flat SiC monolayer.
In contrast, the density of conduction states near the Fermi energy is generated by a hybridization of the $p_z$-orbitals of Si and C atoms where the $p_z$-orbital of the Si atoms is much higher.
With increasing $\Delta$, the conduction density of states of the $s$- and the $p$-orbital for
both the Si and the C atoms moves towards the Fermi energy resulting in a reduction of
the semiconducting energy band gap.
In addition, the contribution of the $s$- and $p_{x,y}$-orbitals of both the Si and the C atoms to the density of states and thus the valence and conduction bands are found with increasing value of $\Delta$.
This is an indication that the planar buckling influences the $\sigma$ and the $\pi$ bonds of a SiC
monolayer leading to enhanced $\sigma\text{-}\pi$ common bonds in the density of states as well as the band structure.

\begin{figure}[htb]
	\centering
	\includegraphics[width=0.22\textwidth]{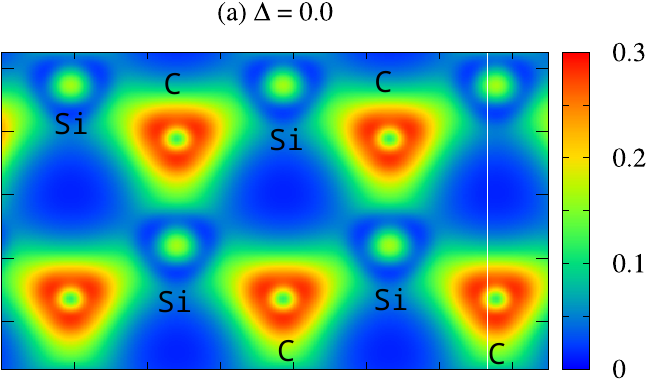}\\
	\includegraphics[width=0.22\textwidth]{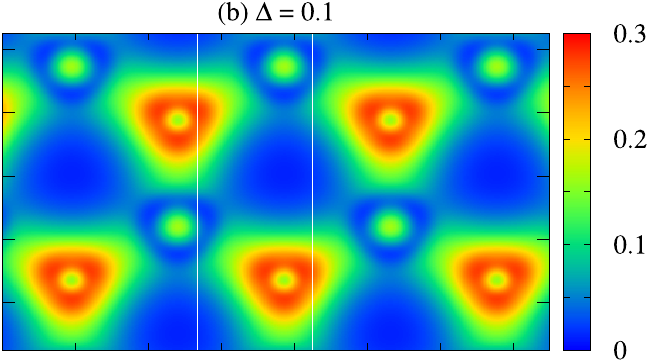}
	\includegraphics[width=0.22\textwidth]{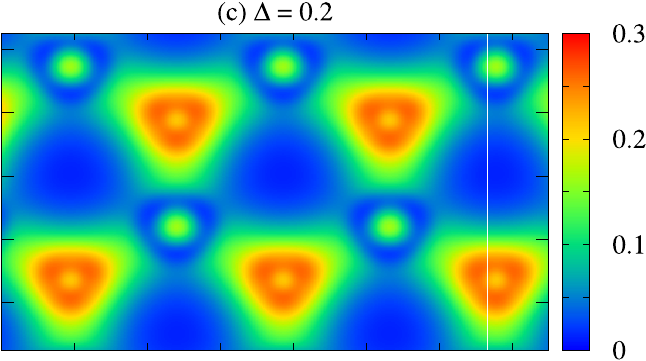}\\
	\includegraphics[width=0.22\textwidth]{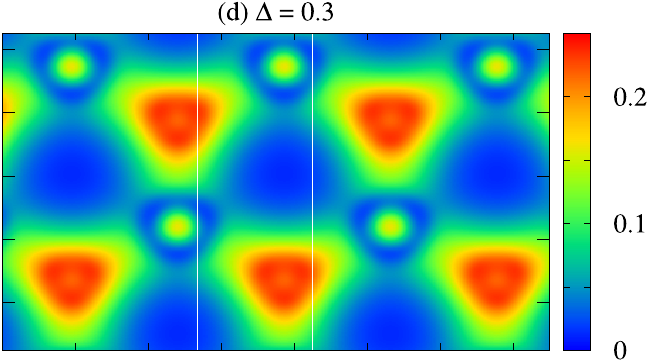}
	\includegraphics[width=0.22\textwidth]{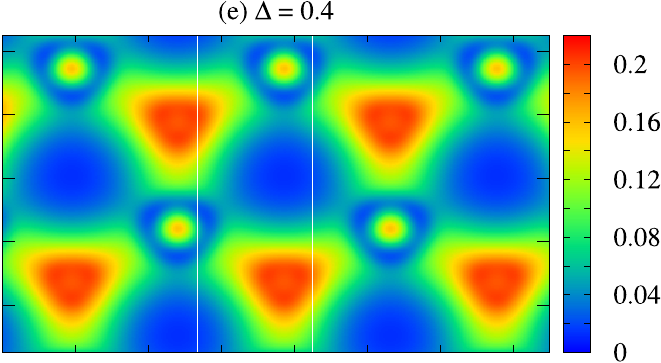}\\
	\includegraphics[width=0.22\textwidth]{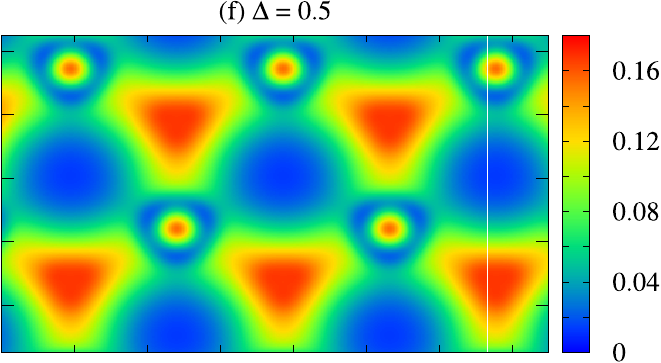}
	\includegraphics[width=0.22\textwidth]{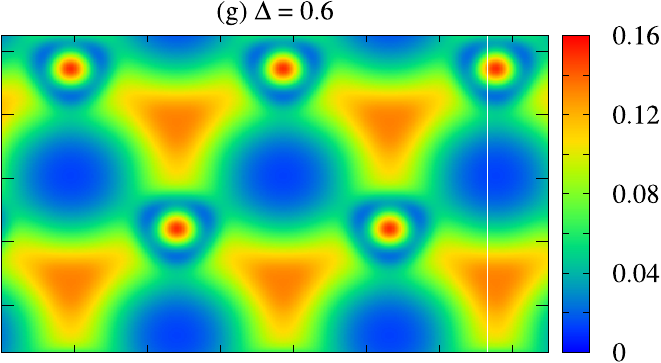}
	\caption{Electron density of SiC monolayers with planar buckling, $\Delta = 0.0$ (a), $0.1$ (b), $0.2$ (c), $0.3$ (d), $0.4$ (e), $0.5$ (f), and $0.6$~$\angstrom$ (g).}
	\label{fig05}
\end{figure}

The electron density distribution describing the valence electron density of a SiC monolayer with different values of $\Delta$ is presented in \fig{fig05} \cite{ABDULLAH2022106409}. The Si and the C atoms have $3s^2 3p^2$ and $2s^2 2p^2$ valence electrons, respectively.
Although the Si and the C atoms have the same number of valence electrons, the distribution of the valence electrons around the Si and the C atoms in a flat SiC monolayer are quite different from each other.
The electron density around Si is localized on the atom, whereas that around C is spatially spread. It corresponds to the difference in electronegativities between the C and the Si atoms, where the C atoms are more electronegative than the Si atoms. When the planar buckling increased, the planes of the Si and the C atoms are slightly separated leading to almost the same electron distribution around the Si and the C atoms, but the distribution of the electrons localized around the atoms slighly change their shape (see \fig{fig05}g). The contributions to the density of states were also modified as a result of the buckling effects. In a flat SiC monolayer, the C atoms used to have a larger charge density than the Si atoms due to their higher electronegativity, while this contribution is altered due to the planar buckling caused by the differences in the bond lengths, which are elongated with $\Delta$. (see \fig{fig05}g).

\subsection{Mechanical responses}

\begin{figure}[htb]
	\centering
	\includegraphics[width=0.4\textwidth]{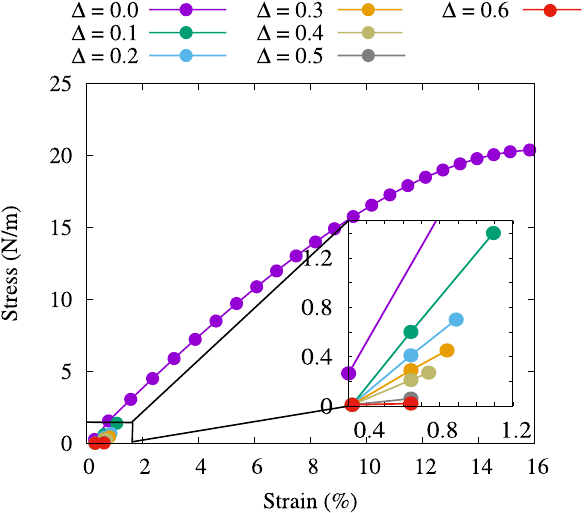}
	\caption{Stress strain curve of the SiC monolayer with buckling parameter, $\Delta = 0.0$ (purple), $0.1$ (green), $0.2$ (light blue), $0.3$ (orange), $0.4$ (olive), $0.4$ (light brown), $0.5$ (gray), and $0.6$~$\angstrom$ (red) in the case of E$_{\parallel}$ (a), and E$_{\perp}$ (b). The vertical black and red lines show different regions of the electromagnetic spectrum.}
	\label{fig06}
\end{figure}

Based on the investigation of stable structures, the mechanical response under uniaxial strain along either the armchair or zigzag direction can be calculated.
The buckling parameter influences the Si-C bond length, and the modifications of bond lengths affect
the mechanical properties of the SiC monolayer. The stress-strain curves of the flat and buckled SiC monolayers are displayed in \fig{fig06} for the Zigzag direction of the structure. The mechanical response along the armchair direction is very similar to the one in the zigzag direction due to the symmetric properties of the SiC monolayer in both the $x$- and the $y$-direction, we thus provide only the stress-strain curve along the zigzag direction.

Our stress-strain curve for flat SiC monolayer is consistent with those existing in the literature, indicating the reliability of the present computational scheme \cite{Lu_2018}.
Similar to previous DFT studies on 2D materials \cite{C5CP02412A}, all monolayer structures with
different stoichiometry also obey linear relationships with strain within a small strain range, and deviate from the linear elasticity under large deformation.
The ultimate tensile strength for a flat SiC is found to be $20.38$ N/m, and the fracture point is $15.8\%$. The fracture point can be defined as the point of strain where the monolayer separates.
At the fracture point, the strain approaches its maximum value, and the monolayer fractures, even though the corresponding stress may be less than the ultimate strength at this point.

In the presence of planar buckling, both the ultimate tensile strength and the values for the fracture point are extremely decreased (inset) in such away that the ultimate tensile strength and the fracture point are inversely proportional to the planar buckling. This is caused by the fact that the planar buckling increases the Si-C bond length, and longer bond length will break easier. Consequently, the buckled SiC monolayer has a lower mechanical response under a uniaxial strain.

\subsection{Thermal properties}

\begin{figure}[htb]
	\centering
	\includegraphics[width=0.45\textwidth]{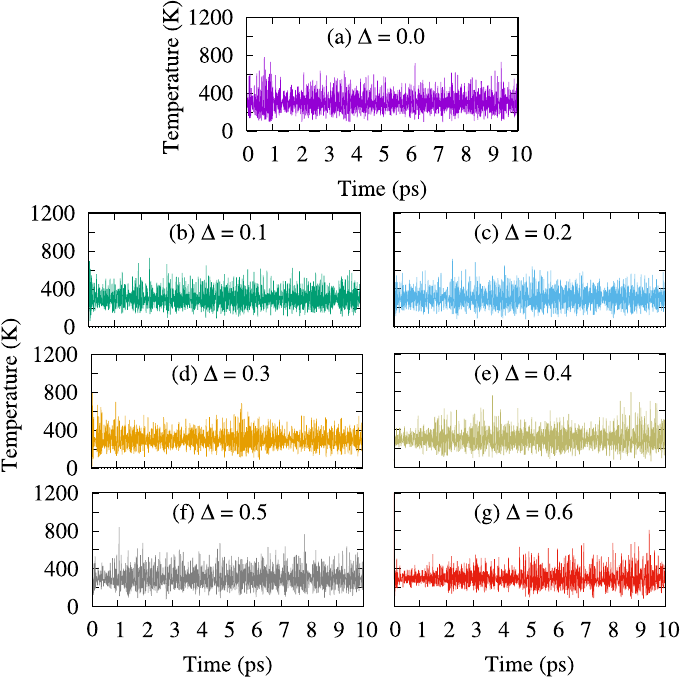}
	\caption{Temperature versus the AIMD simulation time steps at $300$ K for optimized SiC monolayers with planar buckling, $\Delta = 0.0$ (a), $0.1$ (b), $0.2$ (c), $0.3$ (d), $0.4$ (e), $0.5$ (f), and $0.6$~$\angstrom$ (g).}
	\label{fig07}
\end{figure}

In this section, thermal properties such as the heat capacity are presented in addition to the temperature-time curve for the flat and planar SiC monolayers.
The thermal stability is calculated up to 10 ps with a time step of $1.0$~fs as is demonstrated
in \fig{fig07}. The temperature curve of the pure and the buckled SiC monolayers neither displays large fluctuations in the temperature nor serious structure disruptions or bond breaking at $300$~K. This indicates that the pure and the buckled SiC monolayers are thermodynamically stable nanosheets.

\begin{figure}[htb]
	\centering
	\includegraphics[width=0.4\textwidth]{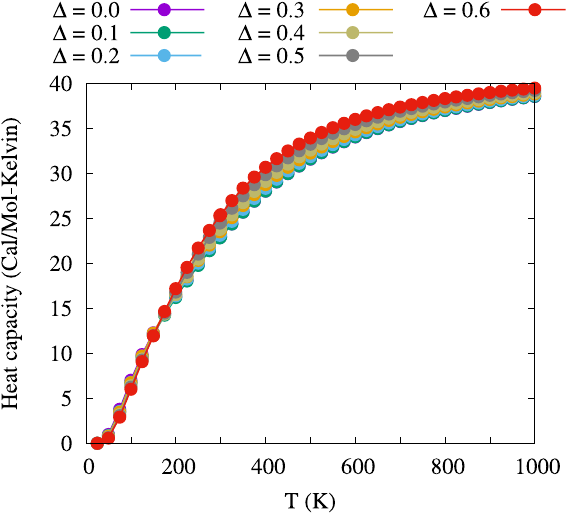}
	\caption{Heat capacity of the SiC monolayer with buckling parameter, $\Delta = 0.0$ (purple), $0.1$ (green), $0.2$ (light blue), $0.3$ (orange), $0.4$ (olive), $0.4$ (light brown), $0.5$ (gray), and $0.6$~$\angstrom$ (red).}
	\label{fig08}
\end{figure}

The heat capacity defines the ratio of the heat absorbed by a material to the temperature change. The heat capacity of a flat and a buckled SiC are shown in \fig{fig08}. The heat
capacity increases with temperature, but becomes almost constant at high values of temperature, $T > 700$ K. It is expected to have higher heat capacity for the systems with stronger bonds,
which is a predicted trend for the heat capacity consistent with the classical theory \cite{MORTAZAVI2021100257}. This means that a higher heat capacity can be obtained from a stronger bond structure. In fact, a bond becomes stronger if the electronegativity difference across the bond increases.
In our buckled monolayers, the buckling does not much affect the electronegativity difference across a bond, and the strength of the bonds is thus almost unchanged with increasing buckling parameter. Consequently, the heat capacity is slightly changed with the buckling parameter as shown in \fig{fig08}.
This indicates that the thermal properties such as heat capacity are not sensitive to the buckling parameter.

\subsection{Optical properties}

\begin{figure*}[htb]
	\centering
	\includegraphics[width=0.9\textwidth]{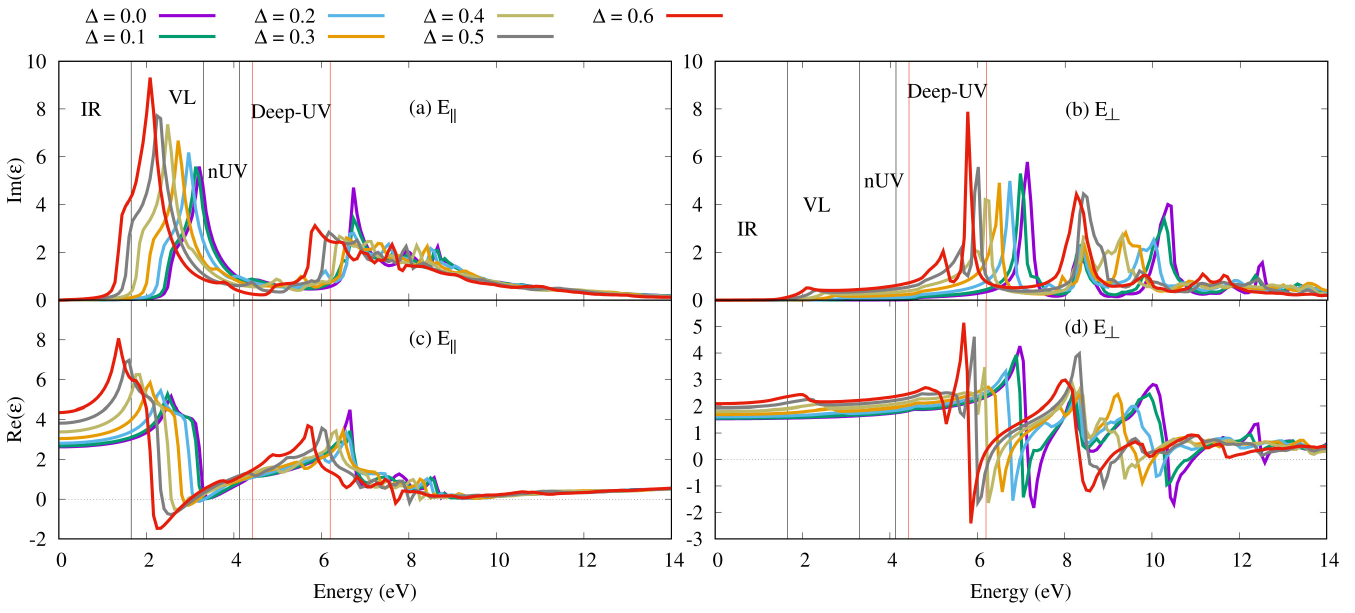}
	\caption{Imaginary, Im($\varepsilon$), (a,b) and real, Re($\varepsilon$), (c,d) parts of dielectric function for the SiC monolayer  with buckling parameter, $\Delta = 0.0$ (purple), $0.1$ (green), $0.2$ (light blue), $0.3$ (orange), $0.4$ (olive), $0.5$ (gray), and $0.6$~$\angstrom$ (red) in the case of E$_{\parallel}$ (left panel), and E$_{\perp}$ (right panel). The vertical black and red lines show different regions of electromagnetic wave.}
	\label{fig09}
\end{figure*}

Next, the optical properties of a SiC monolayer in relation to the planar buckling are investigated. The dielectric function, the absorption coefficient, and the optical conductivity are all calculated using the random phase approximation (RPA) \cite{ren2012random}. In order to achieve great accuracy, the RPA in the QE package employs a dense $100 \times 100 \times 1$ mesh grid in the Brillouin zone \cite{PhysRev.115.786}. For parallel (E$_{\parallel}$) and perpendicular (E$_{\perp}$) polarization of the incoming electric field, the imaginary, Im($\varepsilon$), and the real, Re($\varepsilon$), components of the dielectric function are shown in the \fig{fig09}.

In the case of E$_{\parallel}$, a significant peak in Im($\varepsilon$) for flat SiC monolayer starts to be produced at $2.5$~eV and it reaches maximum value at $3.2$~eV in the visible close to the near-UV spectrum as is presented in \fig{fig09}a (purple line). This is in accordance with previous findings \cite{majidi2018optical} for a stable planar structure.
The result indicated that this value is associated with the SiC's electronic band gap, confirming that SiC is a semiconductor.

As $\Delta$ is increased, the peak moves to lower energy, when the incoming electric field is polarized in parallel (E$_{\parallel}$). The major peak, which corresponds to the optical band gap, has been shifted down to the deep visible and the near-IR areas. The shift of the peak implies the appearance of additional states close to the Fermi energy. This finding is advantageous for optoelectronic devices that work with visible light. On the other hand, as a result of the change in the band structure, as well as the emergence of a shift in the band gap's direction, the peak's intensity is increased. The illustrated electromagnetic spectrum regimes for Im($\varepsilon$) show a considerable tuning to lower energy with increasing $\Delta$ for the case of (E$_{\perp}$).

\lipsum[0]

The real component of the dielectric function is associated to a variety of fascinating physical phenomena, such as the static dielectric constant and the plasmon energy of the SiC monolayer.
The real component of the dielectric function at zero energy, also known as the static dielectric constant Re$(\varepsilon(0))$, is another significant quantity to consider.
Because the value of Re$(\varepsilon(\omega))$ is inversely related to the band gap, Re$(\varepsilon(\omega)) \approx 1/E_{g}$, the value of Re$(\varepsilon(0))$ increases with  $\Delta$ \cite{PhysRev.128.2093}. With increasing $\Delta$, the band gap is narrowed, and the value of Re$(\varepsilon(0))$ rises. Re$(\varepsilon)$ increases faster for E$_{\parallel}$ than for E$_{\perp}$, showing that Re$(\varepsilon)$ exhibits an anisotropy regarding the polarization of the incoming field.

Another interesting phenomena are the plasmons which are connected to both the imaginary and the real components of the dielectric function. The plasmons are a group of valence or conduction electrons in a material participating in collective oscillations. The real component of a material's complex dielectric function depicts the electromagnetic wave transmission across the medium, whereas the imaginary component describes single particle excitations, which are characterized by interband transitions \cite{raether2006excitation, egerton2011introduction}.
At a plasmon energy, the real part of dielectric function is zero, while the imaginary part has a maximum value. So, the real component of the dielectric function changes its sign from positive to negative at the plasmon frequency, as well as at the plasmon energy.
In the case of E$_{\parallel}$, the plasmon energy of a flat SiC monolayer stays unchanged and does not pass the negative sign. It indicates that the flat SiC monolayer does not support plasmonic oscillations in the selected energy range of the applied electromagnetic wave.
When $\Delta$ grows, the plasmon energy appears at $\Delta = 0.2$~$\angstrom$ and
the plasmon energy shifts to a lower energy scale, indicated by the real part of the dielectric function crossing the zero point assuming negative values, indicating that the electrons in a SiC monolayer exhibit plasmonic behavior with increasing $\Delta$, as is shown in \fig{fig09}. The same mechanism can be applied to the plasmonic behavior when E$_{\perp}$ is considered.

\begin{figure}[htb]
	\centering
	\includegraphics[width=0.45\textwidth]{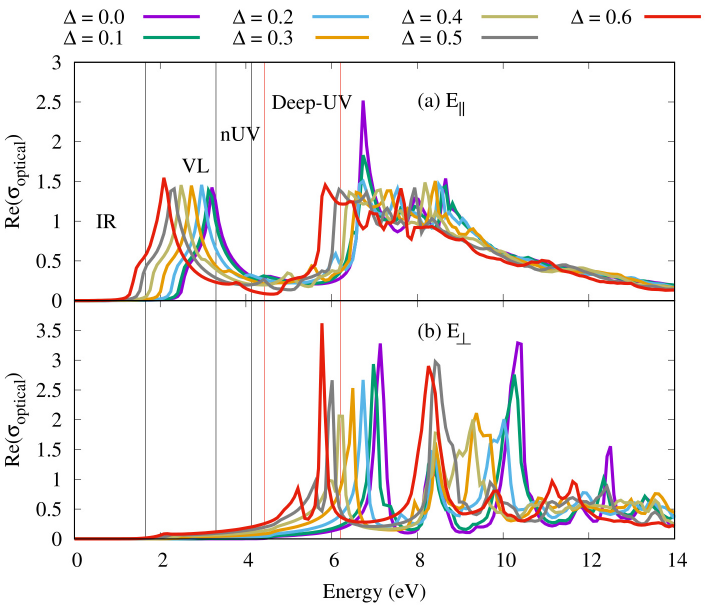}
	\caption{Optical conductivity (real part) of the SiC monolayer with buckling parameter, $\Delta = 0.0$ (purple), $0.1$ (green), $0.2$ (light blue), $0.3$ (orange), $0.4$ (olive), $0.4$ (light brown), $0.5$ (gray), and $0.6$~$\angstrom$ (red) in the case of E$_{\parallel}$ (a), and E$_{\perp}$ (b). The vertical black and red lines show different regions of the electromagnetic spectrum.}
	\label{fig10}
\end{figure}

Next, the information on the electronic band structures in combination with the incoming field E$_{\parallel}$ (a) and E$_{\perp}$ (b) are used to determine the actual components of the optical conductivity spectra for buckled and planar SiC monolayers. For SiC monolayers with planar structures below $2.5$ eV (E$_{\parallel}$) and $4.2$ eV (E$_{\perp}$), $\sigma_{\rm optical}$ is 0 as is shown in \fig{fig10}, and the change follows the same pattern as the change in the dielectric function.
The first intense peak of flat SiC monolayer in $\sigma_{\rm optical}$ for E$_{\parallel}$ and E$_{\perp}$ is found at $3.2$ ($\sigma_{\rm optical} = 1.42$) and $7.13$~eV ($\sigma_{\rm optical} = 3.3$), respectively, which is in a good agreement with previous studies of flat SiC monolayer \cite{HOUMAD20161867}, when noticing that the value of the optical conductivity of our SiC monolayer is smaller due to different size of our supercell.

The visible region of E$_{\parallel}$ exhibits a strong peak, and the intensity of the maximum is slightly increased as $\Delta$ increases, but its position is dislocated to lower energy, revealing the optical band gap.
In addition, there is a significant peak in the UV region at high photon energy $6\text{-}8$~eV, which corresponds to the transitions of to higher energy levels. The peak's intensity is quite high for flat SiC monolayers in this area, but it is decreased by $\Delta$ and moves to the deep visible region, a new conduction mechanism arises. Moreover, when planar buckling is intensified, the polarization direction of the incoming light has a major impact on the optical conductivity, in the case of of E$_{\perp}$. Increased planar buckling alters the structure of the spectra, allowing direct transitions, while maintaining the crystal momentum without a significant change in the wave vector.

\begin{figure}[htb]
	\centering
	\includegraphics[width=0.45\textwidth]{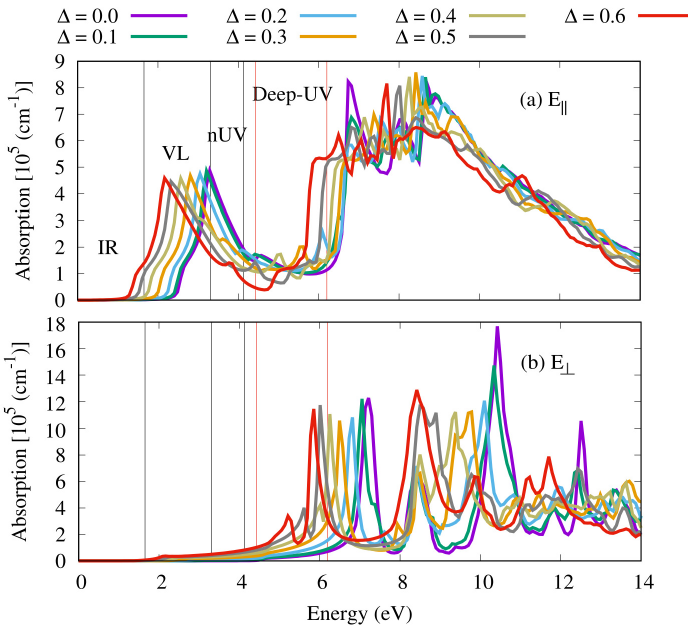}
	\caption{Absorption coefficient of the SiC monolayer  with buckling parameter, $\Delta = 0.0$ (purple), $0.1$ (green), $0.2$ (light blue), $0.3$ (orange),  $0.4$ (olive), $0.5$ (gray) and $0.6$~$\angstrom$ (red) in the case of E$_{\parallel}$ (a), and E$_{\perp}$ (b). The vertical black and red lines show different regions of the electromagnetic spectrum.}
	\label{fig11}
\end{figure}

Our final goal is the absorption spectra presented in \fig{fig11}, which displays the absorption coefficient of a SiC monolayer versus energy for both polarizations in a flat structure and layers with different levels of planar buckling. In contrast to the optical conductivity, intensity of the main peak in the absorption spectra located around $2.0\text{-}4.0$~eV for E$_{\parallel}$ is slightly decreased with $\Delta$.

\section{Conclusions}\label{Sec:Conclusion}

The electronic structure and optical properties of a SiC monolayer are studied using density functional theory calculation for a varying degree of the planar buckling. With an increase in $\Delta$, the band structure changes and the band gap is narrowed. The contributions to the density of states are modified as a result of the buckling effects. In a flat SiC monolayer, the C atoms have a larger charge density than the the Si atoms due to their higher electronegativity, however this contribution varies due to planar buckling causing changes in the bond lengths. Parallel and perpendicular polarization of incoming electric fields with respect to the plane of the SiC monolayer are employed to explore the optical characteristics of the SiC monolayer. With increasing buckling strength, the optical spectra are enhanced for both polarizations. There are shifts for the two main peaks of the real and the imaginary parts of the dielectric function changing from the Deep-UV to the near visible area owing to the appearance of $\sigma\text{-}\pi$ bonds in the sp$^3$ hybridization. Small peaks are related to the electronic band gap and the system remains a semiconductor as the band gap narrows, but strong peaks are associated with interband transitions owing to the frequency dependence of the real component of conductivity.  As a result, the optical characteristics can be tuned by changing the planar buckling in the visible region.
A flat SiC monolayer exhibits good optical properties in the near visible area, whereas the absorption region is changed by increasing planar buckling from the near to the far visible region, resulting in a stronger $\sigma\text{-}\pi$ bond. The SiC siligraphene is a flexible material that can be used in the applications of smart nanomaterial in which the findings are remarkable.

\section{Acknowledgment}
The University of Sulaimani and the Research Center of Komar University of Science and Technology provided financial assistance for this project.
The calculations were carried out using resources given by the University of Sulaimani's Division of Computational Nanoscience.



\begin{thebibliography}{10}
	\expandafter\ifx\csname url\endcsname\relax
	\def\url#1{\texttt{#1}}\fi
	\expandafter\ifx\csname urlprefix\endcsname\relax\def\urlprefix{URL }\fi
	\expandafter\ifx\csname href\endcsname\relax
	\def\href#1#2{#2} \def\path#1{#1}\fi
	
	\bibitem{gupta2015recent}
	A.~Gupta, T.~Sakthivel, S.~Seal, Recent development in 2D materials beyond
	graphene, Progress in Materials Science 73 (2015) 44--126.
	
	\bibitem{peng2015two}
	B.~Peng, P.~K. Ang, K.~P. Loh, Two-dimensional dichalcogenides for
	light-harvesting applications, Nano Today 10~(2) (2015) 128--137.
	
	\bibitem{abdullah2021properties}
	N.~R. Abdullah, B.~J. Abdullah, C.-S. Tang, V.~Gudmundsson, Properties of $BC_6N$
	monolayer derived by first-principle computation: Influences of interactions
	between dopant atoms on thermoelectric and optical properties, Materials
	Science in Semiconductor Processing 135 (2021) 106073.
	
	\bibitem{novoselov2004electric}
	K.~S. Novoselov, A.~K. Geim, S.~V. Morozov, D.-e. Jiang, Y.~Zhang, S.~V.
	Dubonos, I.~V. Grigorieva, A.~A. Firsov, Electric field effect in atomically
	thin carbon films, Science 306~(5696) (2004) 666--669.
	
	\bibitem{mak2016photonics}
	K.~F. Mak, J.~Shan, Photonics and optoelectronics of 2D semiconductor
	transition metal dichalcogenides, Nature Photonics 10~(4) (2016) 216--226.
	
	\bibitem{vogt2012silicene}
	P.~Vogt, P.~De~Padova, C.~Quaresima, J.~Avila, E.~Frantzeskakis, M.~C. Asensio,
	A.~Resta, B.~Ealet, G.~Le~Lay, Silicene: Compelling experimental evidence for
	graphenelike two-dimensional silicon, Physical review letters 108~(15) (2012)
	155501.
	
	\bibitem{mannix2018borophene}
	A.~J. Mannix, Z.~Zhang, N.~P. Guisinger, B.~I. Yakobson, M.~C. Hersam,
	Borophene as a prototype for synthetic 2D materials development, Nature
	nanotechnology 13~(6) (2018) 444--450.
	
	\bibitem{zhu2015epitaxial}
	F.-f. Zhu, W.-j. Chen, Y.~Xu, C.-l. Gao, D.-d. Guan, C.-h. Liu, D.~Qian, S.-C.
	Zhang, J.-f. Jia, Epitaxial growth of two-dimensional stanene, Nature
	materials 14~(10) (2015) 1020--1025.
	
	\bibitem{zheng2019adsorption}
	F.~Zheng, H.~Dong, Y.~Ji, Y.~Li, Adsorption of hydrazine on $XC_3$ (X= B, Al, N,
	Si, and Ge) nanosheets: A computational study, International Journal of
	Hydrogen Energy 44~(12) (2019) 6055--6064.
	
	\bibitem{anikina2020elucidating}
	E.~Anikina, T.~Hussain, V.~Beskachko, R.~Ahuja, Elucidating hydrogen storage
	properties of two-dimensional siligraphene (SiC$_8$) monolayers upon selected
	metal decoration, Sustainable Energy \& Fuels 4~(11) (2020) 5578--5587.
	
	\bibitem{houmad2018thermal}
	M.~Houmad, A.~El~Kenz, A.~Benyoussef, Thermal and electrical properties of
	siligraphene and its derivatives, Optik 157 (2018) 936--943.
	
	\bibitem{guan2019computational}
	J.~Guan, L.~Zhang, K.~Deng, Y.~Du, E.~Kan, Computational dissection of 2D $SiC_7$
	monolayer: a direct band gap semiconductor and high power conversion
	efficiency, Advanced Theory and Simulations 2~(8) (2019) 1900058.
	
	\bibitem{houmad2016optical}
	M.~Houmad, H.~Zaari, A.~Benyoussef, A.~El~Kenz, Optical properties of titanium
	and iron doped 3C-SiC behaviors Tb-Mbj, Chinese journal of physics 54~(6)
	(2016) 960--967.
	
	\bibitem{abdullah2020electronic}
	N.~R. Abdullah, G.~A. Mohammed, H.~O. Rashid, V.~Gudmundsson, Electronic,
	thermal, and optical properties of graphene like sicx structures: Significant
	effects of Si atom configurations, Physics Letters A 384~(24) (2020) 126578.
	
	\bibitem{lin2012light}
	S.~Lin, Light-emitting two-dimensional ultrathin silicon carbide, The Journal
	of Physical Chemistry C 116~(6) (2012) 3951--3955.
	
	\bibitem{chabi2016graphene}
	S.~Chabi, H.~Chang, Y.~Xia, Y.~Zhu, From graphene to silicon carbide: ultrathin
	silicon carbide flakes, Nanotechnology 27~(7) (2016) 075602.
	
	\bibitem{chen2019low}
	J.~Chen, N.~Li, Y.~Wei, B.~Han, Y.~Zhang, A low-cost approach to fabricate SiC
	nanosheets by reactive sintering from Si powders and graphite, Journal of
	Alloys and Compounds 788 (2019) 345--351.
	
	\bibitem{chabi2021creation}
	S.~Chabi, Z.~Guler, A.~J. Brearley, A.~D. Benavidez, T.~S. Luk, The creation of
	true two-dimensional silicon carbide, Nanomaterials 11~(7) (2021) 1799.
	
	\bibitem{majid2020first}
	A.~Majid, N.~Rani, S.~U.-D. Khan, Z.~A. Almutairi, First principles study of
	structural, electronic and magnetic properties of transition metals doped SiC
	monolayers for applications in spintronics, Journal of Magnetism and Magnetic
	Materials 503 (2020) 166648.
	
	\bibitem{wan2019photoelectric}
	Y.~Wan-Jun, Q.~Xin-Mao, Z.~Chun-Hong, Z.~Zhong-Zheng, Z.~Shi-Yun, Photoelectric
	properties of La, Ce, Th doped 2D SiC: A first principle study, J. Nanomater.
	Mol. Nanotechnol 2018 (2019) 10.
	
	\bibitem{xu2016controlling}
	Z.~Xu, Y.~Li, Z.~Liu, Controlling electronic and optical properties of layered
	SiC and GeC sheets by strain engineering, Materials \& Design 108 (2016)
	333--342.
	
	\bibitem{ABDULLAH2022115554}
	N.~R. Abdullah, H.~O. Rashid, C.-S. Tang, A.~Manolescu, V.~Gudmundsson,
	\href{https://www.sciencedirect.com/science/article/pii/S0921510721005080}{Controlling
		physical properties of bilayer graphene by stacking orientation caused by
		interaction between B and N dopant atoms}, Materials Science and Engineering:
	B 276 (2022) 115554.
	\newblock \href {https://doi.org/https://doi.org/10.1016/j.mseb.2021.115554}
	{\path{doi:https://doi.org/10.1016/j.mseb.2021.115554}}.
	\newline\urlprefix\url{https://www.sciencedirect.com/science/article/pii/S0921510721005080}
	
	\bibitem{belarouci2018two}
	S.~Belarouci, T.~Ouahrani, N.~Benabdallah, A.~Morales-Garcia, I.~Belabbas,
	Two-dimensional silicon carbide structure under uniaxial strains, electronic
	and bonding analysis, Computational Materials Science 151 (2018) 288--295.
	
	\bibitem{zhao2021enhancement}
	Z.~Zhao, Y.~Yong, R.~Gao, S.~Hu, Q.~Zhou, Y.~Kuang, Enhancement of Nitride-GaS
	sensing performance of sic7 monolayer induced by external electric field,
	Vacuum 191 (2021) 110393.
	
	\bibitem{delavari2018electronic}
	N.~Delavari, M.~Jafari, Electronic and optical properties of hydrogenated
	silicon carbide nanosheets: A DFT study, Solid State Communications 275
	(2018) 1--7.
	
	\bibitem{jalilian2016buckling}
	J.~Jalilian, M.~Safari, S.~Naderizadeh, Buckling effects on electronic and
	optical properties of BeO monolayer: First principles study, Computational
	Materials Science 117 (2016) 120--126.
	
	\bibitem{abdullah2022enhanced}
	N.~R. Abdullah, B.~J. Abdullah, C.-S. Tang, V.~Gudmundsson, Enhanced
	ultraviolet absorption in BN monolayers caused by tunable buckling, arXiv
	preprint arXiv:2201.00116 (2022).
	
	\bibitem{PhysRevB.54.11169}
	G.~Kresse, J.~Furthm\"uller,
	\href{https://link.aps.org/doi/10.1103/PhysRevB.54.11169}{Efficient iterative
		schemes for ab initio total-energy calculations using a plane-wave basis
		set}, Phys. Rev. B 54 (1996) 11169--11186.
	\newblock \href {https://doi.org/10.1103/PhysRevB.54.11169}
	{\path{doi:10.1103/PhysRevB.54.11169}}.
	\newline\urlprefix\url{https://link.aps.org/doi/10.1103/PhysRevB.54.11169}
	
	\bibitem{Giannozzi_2009}
	P.~Giannozzi, S.~Baroni, N.~Bonini, M.~Calandra, R.~Car, C.~Cavazzoni,
	D.~Ceresoli, G.~L. Chiarotti, M.~Cococcioni, I.~Dabo, A.~D. Corso,
	S.~de~Gironcoli, S.~Fabris, G.~Fratesi, R.~Gebauer, U.~Gerstmann,
	C.~Gougoussis, A.~Kokalj, M.~Lazzeri, L.~Martin-Samos, N.~Marzari, F.~Mauri,
	R.~Mazzarello, S.~Paolini, A.~Pasquarello, L.~Paulatto, C.~Sbraccia,
	S.~Scandolo, G.~Sclauzero, A.~P. Seitsonen, A.~Smogunov, P.~Umari, R.~M.
	Wentzcovitch,
	\href{https://doi.org/10.1088%2F0953-8984%2F21%2F39%2F395502}{{QUANTUM}
		{ESPRESSO}: a modular and open-source software project for quantum
		simulations of materials}, Journal of Physics: Condensed Matter 21~(39)
	(2009) 395502.
	\newblock \href {https://doi.org/10.1088/0953-8984/21/39/395502}
	{\path{doi:10.1088/0953-8984/21/39/395502}}.
	\newline\urlprefix\url{https://doi.org/10.1088%2F0953-8984%2F21%2F39%2F395502}
	
	\bibitem{giannozzi2017advanced}
	P.~Giannozzi, O.~Andreussi, T.~Brumme, O.~Bunau, M.~B. Nardelli, M.~Calandra,
	R.~Car, C.~Cavazzoni, D.~Ceresoli, M.~Cococcioni, et~al., Advanced
	capabilities for materials modelling with quantum espresso, Journal of
	Physics: Condensed Matter 29~(46) (2017) 465901.
	
	\bibitem{doi:10.1063/1.463940}
	G.~J. Martyna, M.~L. Klein, M.~Tuckerman,
	\href{https://doi.org/10.1063/1.463940}{Nosé–hoover chains: The canonical
		ensemble via continuous dynamics}, The Journal of Chemical Physics 97~(4)
	(1992) 2635--2643.
	\newblock \href {http://arxiv.org/abs/https://doi.org/10.1063/1.463940}
	{\path{arXiv:https://doi.org/10.1063/1.463940}}, \href
	{https://doi.org/10.1063/1.463940} {\path{doi:10.1063/1.463940}}.
	\newline\urlprefix\url{https://doi.org/10.1063/1.463940}
	
	\bibitem{ABDULLAH2021114644}
	N.~R. Abdullah, M.~T. Kareem, H.~O. Rashid, A.~Manolescu, V.~Gudmundsson,
	\href{https://www.sciencedirect.com/science/article/pii/S1386947721000266}{Spin-polarised
		DFT modeling of electronic, magnetic, thermal and optical properties of
		silicene doped with transition metals}, Physica E: Low-dimensional Systems
	and Nanostructures 129 (2021) 114644.
	\newblock \href {https://doi.org/https://doi.org/10.1016/j.physe.2021.114644}
	{\path{doi:https://doi.org/10.1016/j.physe.2021.114644}}.
	\newline\urlprefix\url{https://www.sciencedirect.com/science/article/pii/S1386947721000266}
	
	\bibitem{Chabi_2016}
	S.~Chabi, H.~Chang, Y.~Xia, Y.~Zhu,
	\href{https://doi.org/10.1088/0957-4484/27/7/075602}{From graphene to silicon
		carbide: ultrathin silicon carbide flakes}, Nanotechnology 27~(7) (2016)
	075602.
	\newblock \href {https://doi.org/10.1088/0957-4484/27/7/075602}
	{\path{doi:10.1088/0957-4484/27/7/075602}}.
	\newline\urlprefix\url{https://doi.org/10.1088/0957-4484/27/7/075602}
	
	\bibitem{nano10112226}
	S.~Chabi, K.~Kadel,
	\href{https://www.mdpi.com/2079-4991/10/11/2226}{Two-dimensional silicon
		carbide: Emerging direct band gap semiconductor}, Nanomaterials 10~(11)
	(2020).
	\newblock \href {https://doi.org/10.3390/nano10112226}
	{\path{doi:10.3390/nano10112226}}.
	\newline\urlprefix\url{https://www.mdpi.com/2079-4991/10/11/2226}
	
	\bibitem{doi:10.1021/jp210536m}
	S.~S. Lin, \href{https://doi.org/10.1021/jp210536m}{Light-emitting
		two-dimensional ultrathin silicon carbide}, The Journal of Physical Chemistry
	C 116~(6) (2012) 3951--3955.
	\newblock \href {http://arxiv.org/abs/https://doi.org/10.1021/jp210536m}
	{\path{arXiv:https://doi.org/10.1021/jp210536m}}, \href
	{https://doi.org/10.1021/jp210536m} {\path{doi:10.1021/jp210536m}}.
	\newline\urlprefix\url{https://doi.org/10.1021/jp210536m}
	
	\bibitem{ABDULLAH2022106943}
	N.~R. Abdullah, B.~J. Abdullah, H.~O. Rashid, C.-S. Tang, V.~Gudmundsson,
	\href{https://www.sciencedirect.com/science/article/pii/S1369800122004772}{Study
		of the buckling effects on the electrical and optical properties of the group
		III-Nitride monolayers}, Materials Science in Semiconductor Processing 150
	(2022) 106943.
	\newblock \href {https://doi.org/https://doi.org/10.1016/j.mssp.2022.106943}
	{\path{doi:https://doi.org/10.1016/j.mssp.2022.106943}}.
	\newline\urlprefix\url{https://www.sciencedirect.com/science/article/pii/S1369800122004772}
	
	\bibitem{SHI2013319}
	Y.-L. Shi, J.-M. Zhang, K.-W. Xu,
	\href{https://www.sciencedirect.com/science/article/pii/S1386947713002622}{Structural
		and electronic properties of SiC nanotubes filled with cu nanowires: A
		first-principles study}, Physica E: Low-dimensional Systems and
	Nanostructures 54 (2013) 319--325.
	\newblock \href {https://doi.org/https://doi.org/10.1016/j.physe.2013.07.021}
	{\path{doi:https://doi.org/10.1016/j.physe.2013.07.021}}.
	\newline\urlprefix\url{https://www.sciencedirect.com/science/article/pii/S1386947713002622}
	
	\bibitem{ABDULLAH2022106835}
	N.~R. Abdullah, B.~J. Abdullah, V.~Gudmundsson,
	\href{https://www.sciencedirect.com/science/article/pii/S1293255822000309}{DFT
		study of tunable electronic, magnetic, thermal, and optical properties of a
	  $Ga_2Si_6$ monolayer}, Solid State Sciences 125 (2022) 106835.
	\newblock \href
	{https://doi.org/https://doi.org/10.1016/j.solidstatesciences.2022.106835}
	{\path{doi:https://doi.org/10.1016/j.solidstatesciences.2022.106835}}.
	\newline\urlprefix\url{https://www.sciencedirect.com/science/article/pii/S1293255822000309}
	
	\bibitem{gacevic:hal-00632224}
	Z.~Gacevic, P.~Lefebvre, F.~Bertram, G.~Schmidt, P.~Veit, J.~Christen,
	E.~Calleja, \href{https://hal.archives-ouvertes.fr/hal-00632224}{{Growth and
			Characterization of InGaN/GaN Quantum Dots for violet-blue Applications}},
	{9th International Conference on Nitride Semiconductors - ICNS9.}, poster
	(Jul 2011).
	\newline\urlprefix\url{https://hal.archives-ouvertes.fr/hal-00632224}
	
	\bibitem{Alaal_2016}
	N.~Alaal, V.~Loganathan, N.~Medhekar, A.~Shukla,
	\href{https://doi.org/10.1088/0022-3727/49/10/105306}{First principles
		many-body calculations of electronic structure and optical properties of
		{SiC} nanoribbons}, Journal of Physics D: Applied Physics 49~(10) (2016)
	105306.
	\newblock \href {https://doi.org/10.1088/0022-3727/49/10/105306}
	{\path{doi:10.1088/0022-3727/49/10/105306}}.
	\newline\urlprefix\url{https://doi.org/10.1088/0022-3727/49/10/105306}
	
	\bibitem{PhysRevB.84.085404}
	H.~C. Hsueh, G.~Y. Guo, S.~G. Louie,
	\href{https://link.aps.org/doi/10.1103/PhysRevB.84.085404}{Excitonic effects
		in the optical properties of a sic sheet and nanotubes}, Phys. Rev. B 84
	(2011) 085404.
	\newblock \href {https://doi.org/10.1103/PhysRevB.84.085404}
	{\path{doi:10.1103/PhysRevB.84.085404}}.
	\newline\urlprefix\url{https://link.aps.org/doi/10.1103/PhysRevB.84.085404}
	
	\bibitem{PhysRevB.81.075433}
	E.~Bekaroglu, M.~Topsakal, S.~Cahangirov, S.~Ciraci,
	\href{https://link.aps.org/doi/10.1103/PhysRevB.81.075433}{First-principles
		study of defects and adatoms in silicon carbide honeycomb structures}, Phys.
	Rev. B 81 (2010) 075433.
	\newblock \href {https://doi.org/10.1103/PhysRevB.81.075433}
	{\path{doi:10.1103/PhysRevB.81.075433}}.
	\newline\urlprefix\url{https://link.aps.org/doi/10.1103/PhysRevB.81.075433}
	
	\bibitem{PhysRevB.80.155453}
	H.~\ifmmode~\mbox{\c{S}}\else \c{S}\fi{}ahin, S.~Cahangirov, M.~Topsakal,
	E.~Bekaroglu, E.~Akturk, R.~T. Senger, S.~Ciraci,
	\href{https://link.aps.org/doi/10.1103/PhysRevB.80.155453}{Monolayer
		honeycomb structures of group-IV elements and III-V binary compounds:
		First-principles calculations}, Phys. Rev. B 80 (2009) 155453.
	\newblock \href {https://doi.org/10.1103/PhysRevB.80.155453}
	{\path{doi:10.1103/PhysRevB.80.155453}}.
	\newline\urlprefix\url{https://link.aps.org/doi/10.1103/PhysRevB.80.155453}
	
	\bibitem{ABDULLAH2022106409}
	N.~R. Abdullah, B.~J. Abdullah, V.~Gudmundsson,
	\href{https://www.sciencedirect.com/science/article/pii/S136980012100740X}{High
		thermoelectric and optical conductivity driven by the interaction of boron
		and nitrogen dopant atoms with a 2D monolayer beryllium oxide}, Materials
	Science in Semiconductor Processing 141 (2022) 106409.
	\newblock \href {https://doi.org/https://doi.org/10.1016/j.mssp.2021.106409}
	{\path{doi:https://doi.org/10.1016/j.mssp.2021.106409}}.
	\newline\urlprefix\url{https://www.sciencedirect.com/science/article/pii/S136980012100740X}
	
	\bibitem{Lu_2018}
	X.~K. Lu, T.~Y. Xin, Q.~Zhang, Q.~Xu, T.~H. Wei, Y.~X. Wang,
	\href{https://doi.org/10.1088%2F1361-6528%2Faac337}{Versatile mechanical
		properties of novel g-$SiC_x$ monolayers from graphene to silicene: a
		first-principles study}, Nanotechnology 29~(31) (2018) 315701.
	\newblock \href {https://doi.org/10.1088/1361-6528/aac337}
	{\path{doi:10.1088/1361-6528/aac337}}.
	\newline\urlprefix\url{https://doi.org/10.1088%2F1361-6528%2Faac337}
	
	\bibitem{C5CP02412A}
	Z.~G. Fthenakis, N.~N. Lathiotakis,
	\href{http://dx.doi.org/10.1039/C5CP02412A}{Graphene allotropes under extreme
		uniaxial strain: an ab initio theoretical study}, Phys. Chem. Chem. Phys. 17
	(2015) 16418--16427.
	\newblock \href {https://doi.org/10.1039/C5CP02412A}
	{\path{doi:10.1039/C5CP02412A}}.
	\newline\urlprefix\url{http://dx.doi.org/10.1039/C5CP02412A}
	
	\bibitem{MORTAZAVI2021100257}
	B.~Mortazavi, F.~Shojaei, T.~Rabczuk, X.~Zhuang,
	\href{https://www.sciencedirect.com/science/article/pii/S2452262721000362}{High
		tensile strength and thermal conductivity in BeO monolayer: A
		first-principles study}, FlatChem 28 (2021) 100257.
	\newblock \href {https://doi.org/https://doi.org/10.1016/j.flatc.2021.100257}
	{\path{doi:https://doi.org/10.1016/j.flatc.2021.100257}}.
	\newline\urlprefix\url{https://www.sciencedirect.com/science/article/pii/S2452262721000362}
	
	\bibitem{ren2012random}
	X.~Ren, P.~Rinke, C.~Joas, M.~Scheffler, Random-phase approximation and its
	applications in computational chemistry and materials science, Journal of
	Materials Science 47~(21) (2012) 7447--7471.
	
	\bibitem{PhysRev.115.786}
	H.~Ehrenreich, M.~H. Cohen,
	\href{https://link.aps.org/doi/10.1103/PhysRev.115.786}{Self-consistent field
		approach to the many-electron problem}, Phys. Rev. 115 (1959) 786--790.
	\newblock \href {https://doi.org/10.1103/PhysRev.115.786}
	{\path{doi:10.1103/PhysRev.115.786}}.
	\newline\urlprefix\url{https://link.aps.org/doi/10.1103/PhysRev.115.786}
	
	\bibitem{majidi2018optical}
	S.~Majidi, N.~B. Nezafat, D.~Rai, A.~Achour, H.~Ghaziasadi, A.~Sheykhian,
	S.~Solaymani, Optical and electronic properties of pure and fully
	hydrogenated SiC and GeC nanosheets: first-principles study, Optical and
	Quantum Electronics 50~(7) (2018) 1--13.
	
	\bibitem{PhysRev.128.2093}
	D.~R. Penn,
	\href{https://link.aps.org/doi/10.1103/PhysRev.128.2093}{Wave-number-dependent
		dielectric function of semiconductors}, Phys. Rev. 128 (1962) 2093--2097.
	\newblock \href {https://doi.org/10.1103/PhysRev.128.2093}
	{\path{doi:10.1103/PhysRev.128.2093}}.
	\newline\urlprefix\url{https://link.aps.org/doi/10.1103/PhysRev.128.2093}
	
	\bibitem{raether2006excitation}
	H.~Raether, Excitation of plasmons and interband transitions by electrons,
	Vol.~88, Springer, 2006.
	
	\bibitem{egerton2011introduction}
	R.~Egerton, An introduction to eels, in: Electron Energy-Loss Spectroscopy in
	the Electron Microscope, Springer, 2011, pp. 1--28.
	
	\bibitem{HOUMAD20161867}
	M.~Houmad, O.~Dakir, A.~Abbassi, A.~Benyoussef, A.~{El Kenz}, H.~Ez-Zahraouy,
	\href{https://www.sciencedirect.com/science/article/pii/S0030402615016253}{Optical
		properties of sic nanosheet}, Optik 127~(4) (2016) 1867--1870.
	\newblock \href {https://doi.org/https://doi.org/10.1016/j.ijleo.2015.11.017}
	{\path{doi:https://doi.org/10.1016/j.ijleo.2015.11.017}}.
	\newline\urlprefix\url{https://www.sciencedirect.com/science/article/pii/S0030402615016253}
	
\end{thebibliography}

\end{document}